\documentclass{article}
\usepackage{spconf,amsmath,graphicx}
\usepackage{color}
\usepackage{hyperref}
\usepackage{makecell}
\usepackage{booktabs}
\usepackage{tabularx}
\usepackage{multirow}

\title{IQDUBBING: Prosody modeling based on discrete self-supervised speech representation for expressive voice conversion}
\name{\begin{tabular}{c}Wendong Gan$^1$, Bolong Wen$^1$,Ying Yan$^1$, Haitao Chen$^1$, Zhichao Wang$^2$ \\
Hongqiang Du$^2$, Lei Xie$^{2*}$\thanks{*Corresponding author}, Kaixuan Guo$^1$, Hai Li$^1$\end{tabular}}

\address{
  $^1$IQIYI Inc, Chengdu, China\\
  $^2$Audio, Speech and Language Processing Group (ASLP@NPU)\\School of Computer Science,
  Northwestern Polytechnical University, Xi’an, China\\
  }

\begin{document}
\ninept
%
\maketitle
\begin{abstract}
Prosody modeling is important, but still challenging in expressive voice conversion. As prosody is difficult to model, and other factors, e.g., speaker, environment and content, which are entangled with prosody in speech, should be removed in prosody modeling. In this paper, we present \textit{IQDubbing} to solve this problem for expressive voice conversion. To model prosody, we leverage the recent advances in discrete self-supervised speech representation (DSSR). Specifically, prosody vector is first extracted from pre-trained VQ-Wav2Vec model, where rich prosody information is embedded while most speaker and environment information are removed effectively by quantization. To further filter out the redundant information except prosody, such as content and partial speaker information, we propose two kinds of prosody filters to sample prosody from the prosody vector. Experiments show that \textit{IQDubbing} is superior to baseline and comparison systems in terms of speech quality while maintaining prosody consistency and speaker similarity.
\end{abstract}
\section{Introduction}
\label{sec:intro}

Voice conversion (VC)~\cite{ezzine2017comparative} aims to modify a speech signal
uttered by a source speaker to sound as if it is uttered by a
target speaker while retaining the linguistic information. Various approaches have been proposed~\cite{sisman2020overview} for voice conversion. As parallel data~\cite{ezzine2017comparative} is expensive to collect, non-parallel methods~\cite{hsu2016voice,kaneko2018cyclegan,sun2016phonetic} have received significant attention. Among them, phonetic posteriorgram (PPG)~\cite{sun2016phonetic} based method is one of the most popular implementations. PPG is extracted from automatic speech recognition (ASR), which is regarded as speaker independent and can be used to represent linguistic content.  Inspired by PPG, the encoder output vector of an end-to-end ASR model~\cite{liu2021fastsvc} and the  bottleneck features (BN)~\cite{lian2021towards}, have been applied to voice conversion. Recently, discrete representation extracted from self-supervised model VQ-Wav2Vec (VQW2V)~\cite{baevski2019vq} is used to build a voice conversion system~\cite{huang2021any}.

Despite recent progress, modeling prosody from expressive speech~\cite{du2021expressive} for style transfer with voice conversion framework is still a challenging task. Besides linguistic information, transferring the source prosody to the target is vital for many voice conversion tasks, including automatic \textit{dubbing}\footnote{This is how our system IQDubbing named -- IQ is from IQIYI Inc, and dubbing is our target VC application.} for movies in which conversations are emotional in nature. Modeling prosody is not a trivial task; furthermore, it is necessary to remove speaker and content related information from prosody representation. Prosody information can be simply extracted from Mel spectrum (Mel)~\cite{skerry2018towards}. However, the extracted prosody contains speaker information and environment noise~\cite{wang2018style}, especially when the prosody is extracted from character conversations in movies, where speech is inevitably mixed with background speech like music and noise. To solve this problem, adversarial neural network is introduced~\cite{li2021ppg}, while it is hard to optimize. Another way to extract prosody is to use BN as input~\cite{wang2021enriching}, which achieves good results in terms of speech quality. However, the major information embedded in BN is linguistic content information, whose contribution to prosody is limited, resulting in converted speech lacking of expressiveness.

Recent studies~\cite{huang2021any, 2019Unsupervised} confirm that discrete self-supervised speech representation (DSSR) is effective to remove  speaker and environment information, but the speech quality of the converted speech is still affected. In this paper, we leverage the advances of discrete self-supervised speech representation for prosody modeling and aim to further disentangle the prosody information from the discrete representation. In the proposed system, named \textit{IQDubbing}, we first use a \textit{content encoder} to extract the linguistic information from ASR bottleneck features. Second, discrete representation which contains rich prosody information is extracted from a pre-trained VQ-Wav2Vec model. As a result, most of speaker and environment information is removed effectively by quantization. We then design a \textit{prosody encoder} to extract prosody vector from discrete representation. However, it is found that the prosody vector still contains redundant information expect prosody, such as content and partial speaker information. Thus we propose two kinds of \textit{prosody filter} to further sample prosody from the prosody vector. Third, we use a individual speaker encoder to represent the speaker identity. Finally, the decoder takes content, prosody and speaker vectors as input and the output is mel spectrum for target waveform generation. Experiment results show that \textit{IQDubbing} outperforms  baseline and comparison systems in terms of speech quality while maintaining good prosody consistency and speaker similarity.

\begin{figure*}[htb]

\begin{minipage}[b]{1.0\linewidth}
  \centering
  \centerline{\includegraphics[width=17cm]{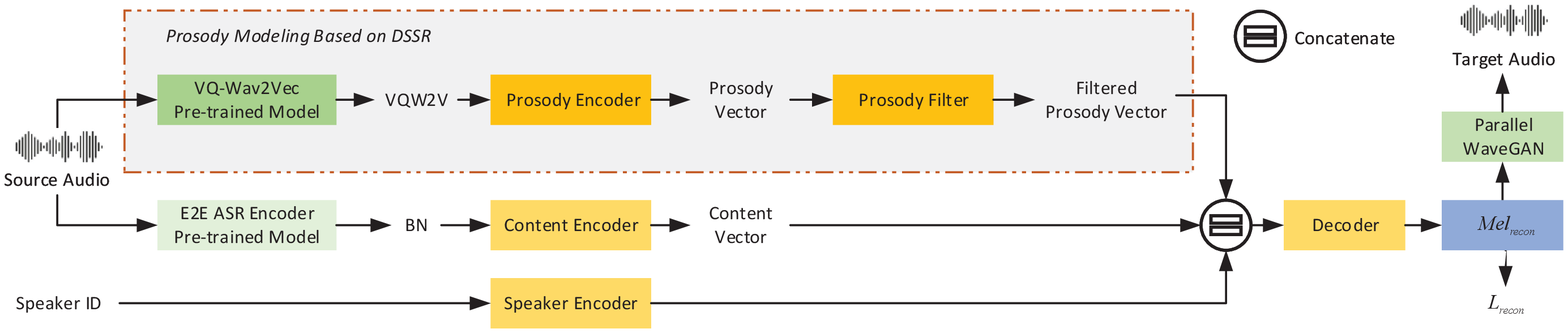}}
\end{minipage}
\caption{Overview of the components of the proposed voice conversion model. First, prosody modeling is based on DSSR. DSSR is discrete self-supervised speech representation. Besides, two kinds of prosody filters: random downsample prosody filter(RDPF) and aligned downsample prosody filter(ADPF), are compared.}
\label{fig:res}
\end{figure*}

The rest of the paper is organized as follows. Section 2 gives an overview on the \textit{IQDubbing} system. The prosody modeling based on DSSR is detailed in Section 3. Section 4 presents the experiments and the comparison of our proposed approach against baseline and comparison systems. Section 5 concludes this paper.

\section{System Overview}
\label{sec:format}

The architecture of IQDubbding is shown in Fig.~1. In general the system follows an encoder-decoder framework, where three individual encoders are adopted, in charge of content extraction, prosody extraction and speaker representation respectively. Specifically for the content extraction, an end-to-end ASR model is first adopted to take source audio as input and its encoder output, or the bottleneck feature (BN), is fed into the content encoder, resulting in a content vector representing the linguistic information. As for the prosody modeling, a pre-trained VQ-Wav2Vec model is adopted to process source audio and output the discrete representation -- VQW2V. The VQW2V is then fed into the prosody encoder, resulting in the prosody vector. To further filter out prosody-unrelated information from the prosody vector, we specifically design a prosody filter to get the filtered prosody vector. The decoder takes content vector, filtered prosody vector and speaker vector as input to reconstruct mel spectrum. Finally, Parallel WaveGAN~\cite{yamamoto2020parallel} is used to synthesize the converted speech.

\section{Prosody Modeling based on DSSR}

Our aim is to transfer not only the linguistic content, but also the style of the source speech to the target speaker. It was previously recognized as a challenging task as linguistic content, speaker identity as well as prosody are entangled in the speech signal. In this paper, we specifically propose a prosody module to output a `pure' prosody vector,  without the linguistic and speaker information, fed into the decoder to result in the converted speech with target speaker's timbre and source speaker's style.

\subsection{VQW2V-based Prosody Encoder}
\label{ssec:subhead}


Recent study adopting VQWav2Vec in VC has shown that the quantization works as a disentanglement operation which helps to remove speaker and environment information, while maintaining rich prosody information from the source speech~\cite{huang2021any, 2019Unsupervised}. Hence based the VQW2V vector outputted by the VQWav2Vec model, we use a prosody encoder to further generate a prosody vector.

\begin{figure}[htb]

\begin{minipage}[b]{1.0\linewidth}
  \centering
  \centerline{\includegraphics[width=8.5cm]{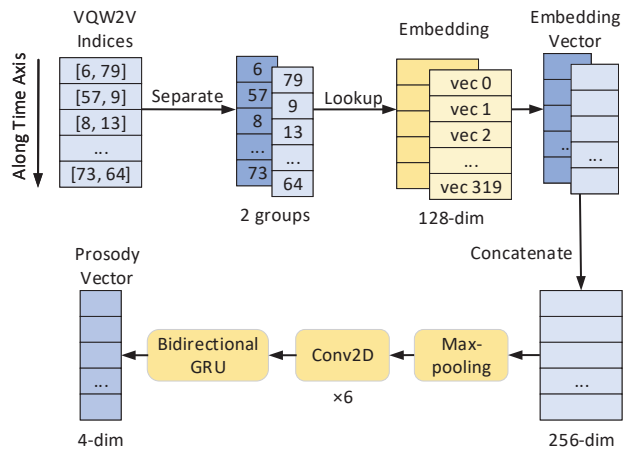}}
\end{minipage}

\caption{VQW2V based prosody encoder. The VQW2V indices are from pre-trained VQ-Wav2Vec model~\cite{2019Unsupervised}.}
\label{fig:res}
\end{figure}


As shown in Fig.2, first the VQW2V prosody encoder takes the indices of VQW2V as input. The codebook size of VQWav2Vec model is set as 320. Second, we divide the indices into two groups along the time and dimension axis. Two groups are used to query from two 128-dim embedding tables to obtain embedding vector. The two queried embedding vectors are concatenated to form frame-wise 256-dim vector. Finally, the  256-dim vector is fed into a max-pooling layer, followed by six 2-D convolution layers and a bidirectional GRU network, which refers to~\cite{li2021ppg}. The  output dimension of VQW2V prosody encoder is set as 4. To our knowledge, this is the first paper to study VQW2V for prosody modeling. We believe that VQW2V is able to improve speech quality and speaker similarity by removing speaker and environment information. 

\subsection{Prosody Filter}
\label{ssec:subhead}


As shown in Fig.1,  we design a module, named as \textit{prosody filter}, to further remove content and speaker information contained in prosody vector. Specifically, we compare two kinds of prosody filter: \textit{random downsample prosody filter} (RDPF) and \textit{aligned downsample prosody filter} (ADPF), which filter out the redundant information by downsamping prosody vector along time axis.

\textbf{Random downsample prosody filter (RDPF)}. Inspired by~\cite{dai2021information},  we design RDPF, which is shown in Fig.3 (a). First, frame level prosody vectors are equally distributed to different phones. A group of vectors corresponds to a phone. Then, we randomly select one prosody vector with a fixed rate $\tau$ from each group as the downsampled vector. Specifically, the downsampled vector is picked out at timestamps $\{\tau-1, 2\tau-1, 3\tau-1, ...\}$. Finally, the downsampled vector is upsampled to form filtered prosody vector, the length of which is the same as prosody vector.

\textbf{Aligned downsample prosody filter (ADPF)}. The second prosody filter ADPF is shown in Fig.~3~(b). First, we use third-party pre-trained model MontrealCorpusTools~\footnote{https://github.com/MontrealCorpusTools/Montreal-Forced-Aligner} to force-align  prosody vectors to the phones to get a time-aligned frame level sequence. Force alignment gives accurate time-aligned sequence.  Then, the prosody vectors are fed into GRU to get corresponding hidden state in frame level. The lengths of hidden states and prosody vectors are same. Therefore, the aligned sequence still suits for the hidden states. Then the last hidden state of each phone is regarded as the downsampled vector. Finally, similar to RDPF, downsampled vector is upsampled to restore the original length of the prosody vector.

Both RDPF and ADPF adopt downsampling technique to remove content and speaker information. Prosody vector is extracted from discrete self-supervised speech representation (DSSR). DSSR is effective to remove  speaker and environment information~\cite{huang2021any, 2019Unsupervised}. Dowmsampling technique compresses prosody vector, which means that a part of the information is lost during the compressing process. As the content and speaker have been represented  by speaker id and content vector, the remaining filtered prosody vector should provide the decoder with sufficient prosody information that is necessary for perfect reconstruction~\cite{qian2019autovc}. Moreover, content changes dramatically among frames in one utterance~\cite{chou2019one}, and each frame level prosody vector may contain the corresponding content. However, the downsampled vector is selected from the corresponding prosody vectors of each phone. Therefore, content information can be further removed.

Comparing RDPF with ADPF, downsampling in RDPF is random, which may lead to corruption of prosody. For example, in Fig.~3, the $4^{th}$ frame belongs to the third phone ``AA2", but the upsampled vector after RDPF is from the second phone. Furthermore, the downsampled vector is just one selected prosody vector, which may be unstable at run-time. For example, the selected prosody vector happens to be noisy. In contrast, ADPF overcomes the above two shortcomings and  achieves more accurate frame level prosody information while removing the redundant information.

\begin{figure*}[htb]
\begin{minipage}[b]{1.0\linewidth}
  \centering
  \centerline{\includegraphics[width=17cm]{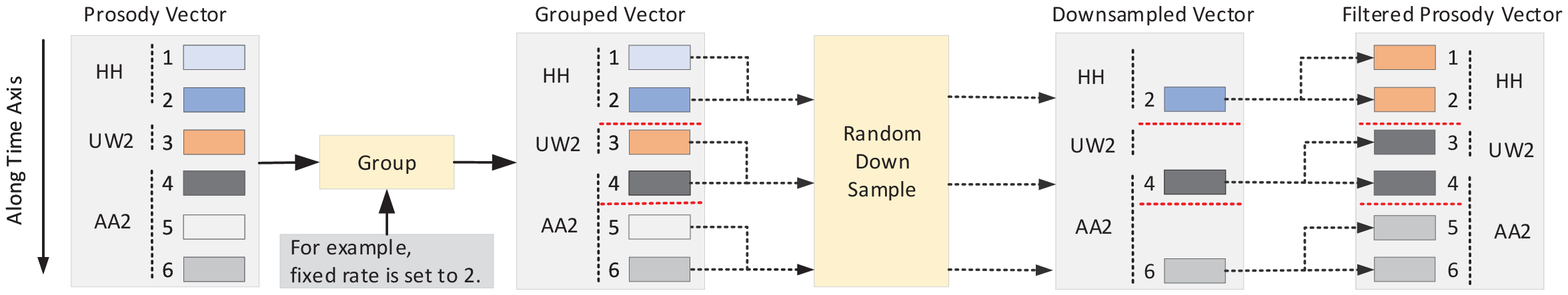}}
  \centerline{(a) Random downsample prosody filter (RDPF). Prosody Vector is grouped by a fixed rate $\tau$. }\medskip
\end{minipage}
\begin{minipage}[b]{1.0\linewidth}
  \centering
  \centerline{\includegraphics[width=17cm]{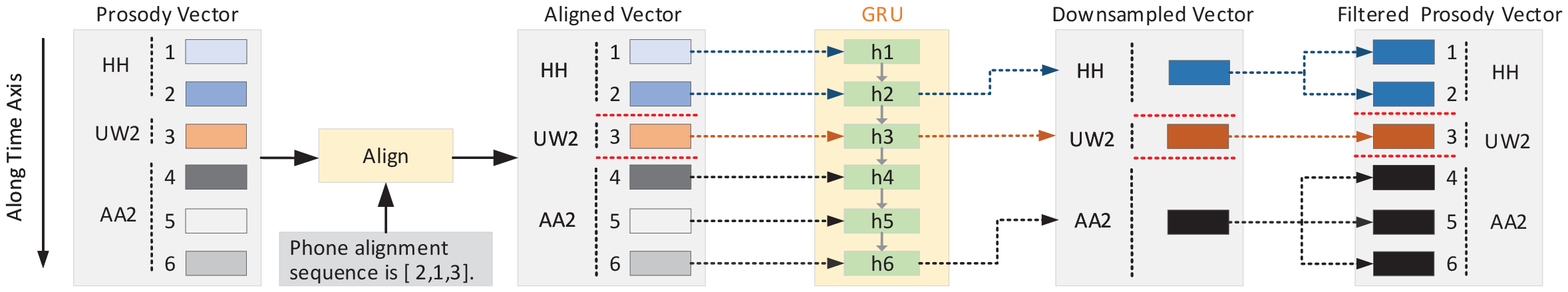}}
  \centerline{(b) Align downsample prosody filter (PLPF). Prosody Vector is aligned by a phone alignment sequence.}\medskip
\end{minipage}

\caption{Prosody filter. Here Mandarin syllable ``hua2" is chosen as an example, which includes 3 phones, ``HH", ``UW2" and ``AA2". The $1^{st}$ and $2^{nd}$ frame belong to ``HH". The $3^{rd}$ frame belongs to ``AA2". The $4^{th}$, $5^{th}$ and $6^{th}$ frame belong to "AA2". Please focus on the filtered prosody vector of each phone.} 
\label{fig:res}
\end{figure*}



\section{EXPERIMENTS}
\label{sec:pagestyle}

\subsection{Dataset and Experimental Setup}
\label{ssec:subhead}

We conduct voice conversion experiments on ESD dataset~\cite{zhou2021emotional}. The ESD dataset covers 5 emotion categories (neutral, happy, angry, sad and surprise), and consists of 350 parallel utterances spoken by 10 native Chinese and 10 native English speakers. For conversion test, 260 utterances are used for evaluation, which contain 10 male speakers and 10 female speakers. The style includes ordinary reading and dubbing speech. All audio files are downsampled to 24kHz. 80 dim Mel is extracted with 50ms frame length and 10ms frame shift.

We trained an end-to-end ASR with 4000 hours mandarin corpus to get 1024-dim BN. The corpus consists of open source dataset\footnote{http://www.openslr.org/68/,  http://www.openslr.org/62/} and internal dataset from TV series which covers different background noises. VQW2V is extracted from third-part pre-trained model~\cite{baevski2019vq}, which is trained with the 960h Librispeech dataset~\cite{panayotov2015librispeech}.  Parallel WaveGAN~\cite{yamamoto2020parallel} is used to reconstruct waveform, which is trained on the ESD dataset.

In order to verify the performance of our proposed approach, we implement the following systems:
\begin{itemize}
\setlength{\itemsep}{0pt}
    \item \textbf{BL}: Baseline system~\cite{sun2016phonetic} is similar to IQDubbing, but is without prosody modeling. Only BN is used as the input of the system.
	\item \textbf{CS}: We implement a comparison system proposed in~\cite{lian2021towards}, which is similar to IQDubbing, but prosody feature is Mel and prosody modeling is without prosody filter. Besides, prosody encoder only includes a max-pooling layer, six 2-D convolution layers and a bidirectional GRU network.
	\item \textbf{IQDubbing-Base}: the proposed system shown in Fig.1,  in which the prosody encoder is VQW2V-based Prosody Encoder introduced in section3.1. Besides, IQDubbing-Base is without prosody filter.  
	\item \textbf{IQDubbing-RDPF}: IQDubbing-Base adopts random downsample prosody filter (RDPF) introduced in Section 3.2. Fixed rate $\tau$ is set as 32, which follows~\cite{dai2021information}. 
	\item \textbf{IQDubbing-ADPF}: IQDubbing-Base adopts aligned downsample prosody filter (ADPF) introduced in Section 3.2.
\end{itemize}


For IQDubbing, we utilize the encoder of Tacotron2~\cite{shen2018natural} as the content encoder and  follows the original configurations. An auto-regressive decoder~\cite{wang2021enriching} is used as our decoder. All the voice conversion models are optimized with Adam optimizer with learning rate decay, which starts from 0.001 and decays every 10 epochs with decay rate 0.7. During the training stage, the models are trained for 140 epochs and batch size is 32.


\subsection{Prosody Consistency Evaluation}
\label{ssec:subhead}

AB test is conducted to verify prosody consistency. All the converted speech are used for listening test. 12  listeners participate in listening test. Listeners are asked to choose the utterance  that is close to source speech in terms of  prosody consistency from paired samples. Higher score indicates that the result is better. The samples can be found here\footnote{https://wblgers.github.io/IQDUBBING-VC.github.io/}.

The results are shown in Fig.4. IQDubbing-Base is significantly better than BL, which verifies that the VQW2V-based prosody encoder has effectively improved the prosody consistency. Besides, IQDubbing-ADPF is better than IQDubbing-Base, IQDubbing-RDPF and CS respectively. The results confirms that IQDubbing-ADPF successfully models the prosody and make the prosody consistent with that of source speech. 

\begin{figure}[!h]
  \centering
  \centerline{\includegraphics[width=8.5cm]{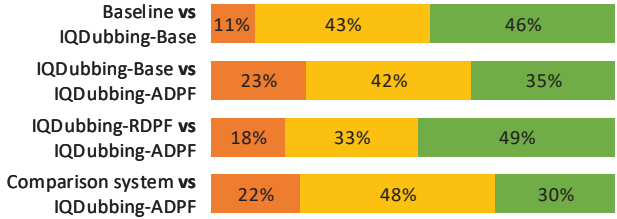}}
\label{fig:res}
\caption{AB test results of prosody consistency.}
\end{figure}


\subsection{Speaker Similarity Evaluation}
\label{ssec:subhead}
In order to verify the performance of each system in terms of speaker similarity, we introduce false acceptance rate of target (FAR) as the indicator of speaker similarity evaluation~\cite{das2020predictions}. The third-party pre-trained speaker verification (SV) system~\footnote{https://github.com/resemble-ai/Resemblyzer} to extract speaker embedding from each utterance~\cite{wan2018generalized}.  We select 140 speakers from Aishell-3~\cite{shi2020aishell} and ESD datasets for evaluation. Ten utterances from each speaker are utilized as the registered utterances to calculate the average speaker embedding.  Higher FAR score indicates that the results of speaker similarity is better.


\begin{table}[!htbp]
\centering
\caption{FAR results of speaker similarity evaluation.}\vspace{5pt}
\begin{tabular}{l|c}
\toprule
System&FAR\\
\midrule
BL&0.833\\
\hline
CS&0.617\\
\hline
IQDubbing-Base&0.818\\
\hline
IQDubbing-RDPF&0.837\\
\hline
IQDubbing-ADPF&\textbf{0.830}\\
\bottomrule
\end{tabular}
\end{table}

The FAR scores of different systems are shown in Table 1. The scores of CS and IQDubbing-Base are lower than BL, which shows that prosody module reduces the performance of speaker similarity as the extracted prosody contains redundant information, such as speaker related information of source speaker. However, the score of IQDubbing-Base is obviously higher than CS, which shows that the VQW2V improves  speaker similarity by quantization. The scores of IQDubbing-RDPF and IQDubbing-ADPF are  higher than IQDubbing-Base and close to BL. The above results show that the prosody filter is effective to improve the speaker similarity.

\subsection{Speech Quality Evaluation}
\label{ssec:subhead}
We conduct  mean opinion score (MOS) tests to evaluate speech quality.  The speech from the target speaker is also used for this evaluation.  Each listener
is asked to give opinion score on a five-point scale (5: excellent, 4: good, 3: fair, 2: poor, 1: bad). 

\begin{table}[!htbp]
\centering
\caption{MOS results of speech quality.}
\vspace{5pt}
\begin{tabular}{l|c}
\toprule
ID&Speech Quality\\
\midrule
Ground Truth& 4.217\\
\hline
BL& 3.733\\
\hline
CS& 3.267\\
\hline
IQDubbing-Base& 3.750\\
\hline
IQDubbing-RDPF& 3.417\\
\hline
IQDubbing-ADPF& \textbf{3.867}\\
\bottomrule
\end{tabular}
\end{table}



The results of MOS tests are shown in Table 2. We find that the score of IQDubbing-Base is higher than BL and CS. However, the score of IQDubbing-RDPF is lower than IQDubbing-Base as RDPF is not stable at runtime. IQDubbing-ADPF achieves the  highest MOS score, which indicates that ADPF helps to improve speech quality by successfully removing content and speaker information in prosody.

\section{CONCLUSION}
\label{sec:typestyle}


Transferring prosody from source to target is vital for expressive voice conversion. However, this is a challenging task as prosody is entangled with other factors including content, speaker and environment information. To solve this problem, we propose \textit{IQDubbing} which models speaker, content and prosody as individual encoders and more importantly, a prosody modeling framework based on discrete self-supervised representation (DSSR) is introduced, aiming to extract rich prosody and remove prosody-unrelated information. Firstly, by using VQ-Wav2Vec for prosody modeling, rich prosody is maintained, while speaker and environment information are removed effectively. And then, since prosody vector is still entangled with redundant information expect prosody, such as content and partial speaker information, we particularly propose two kinds of \textit{prosody filter} to further sample prosody from the prosody vector.  Experiments show that \textit{IQDubbing} is superior to baseline and comparison systems in terms of speech quality and maintains prosody consistency and speaker similarity. Moreover, aligned downsample prosody filter (ADPF) is more effective than random downsample prosody filter (RDPF).

\vfill\pagebreak

\bibliographystyle{IEEEbib}
\small\bibliography{strings,refs}

\end{document}